\documentclass[12pt]{article}
\usepackage{mathtools}
\usepackage{amssymb}
\usepackage{xcolor}
\usepackage{graphicx}
\usepackage{hyperref}
\usepackage{geometry}
\usepackage{enumitem}
\usepackage{setspace}
\usepackage{float}

\usepackage{tikz}
\usetikzlibrary{shapes.callouts, calc}

\newcommand{\mathcalII}{\scalebox{0.7}[1]{\ensuremath{\mathcal{I}\kern-0.33em\mathcal{I}}}}

\hypersetup{hidelinks}
\allowdisplaybreaks[2] 
\title{Macdonald Identities and Exact Formulas for 
Superconformal Indices in Super Yang--Mills Theories}

\makeatletter
\renewcommand\@date{{%
  \vspace{-\baselineskip}%
  \large\centering
  \begin{tabular}{@{}c@{}}
    Yongchao L\"u
  \end{tabular}

  {\small
    School of Physics, Korea Institute for Advanced Study, Seoul 02455, Korea} \\
  \normalsize {lychaoaa@gmail.com}
}}
\makeatother

\begin{document}

\maketitle

\begin{abstract}
We present exact evaluations of superconformal indices for $4d$ $\mathcal{N}{=}1$ and $\mathcal{N}{=}2$ pure Super Yang–Mills theories with arbitrary simple gauge group $G$.
Our approach applies the Macdonald identities for untwisted affine Lie algebras to the integral formulas of the indices, yielding uniform closed formulas valid for all $G$, expressed both as $q$-series and as eta-quotients, related through specialized Macdonald identities.
Using similar techniques, we also derive exact expressions for half Schur indices with Neumann boundary conditions and uncover a bilinear structure of the full Schur index.
Within the framework of holomorphic–topological twists, we further explore connections to the category of line operators, the $K$-theoretical Coulomb branch, Schur quantization, IR formulas for the BPS spectrum, and class $\mathcal{S}$ constructions.
\end{abstract} 

\tableofcontents

\section{Introduction}

In this note, we present closed-form evaluations of the \emph{superconformal}\footnote{We use conventional terminology and a unified formalism applicable to both conformal and non-conformal theories, though the interpretation differs.} index for $4d$ $\mathcal{N}=1$ and $\mathcal{N}=2$ pure supersymmetric Yang–Mills (SYM) theories. Our results provide explicit and uniform expressions for all simple gauge groups $G$.

\paragraph{\underline{Results}} The direct motivation of this work comes from a conjectured closed formula by Okazaki and Smith for the Schur index $\mathbb{I}_{SU(N)}(q)$ for $4d$ $\mathcal{N}=2$ Super Yang-Mill theories with $SU(N)$ gauge group
which is defined by an integration formula following from the UV lagrangian \cite{okazakiS24}
\begin{align}
    \mathbb{I}_{SU(N)}(q) = \frac{(q,q)^{2N-2}}{N!} \oint \frac{dz}{(2\pi i)^{N-1} z} \prod_{i \neq j} \theta(z_i/z_j,q)\,.
\end{align}
where $\theta(x,q)=(x,q)(q/x,q)$ denotes the theta function built from the $q$-Pochhammer symbol $(x,q)=\prod_{k=0}^{\infty}(1-xq^k)$.
The conjectured closed formula admits both product and summation forms\footnote{Similar results for the $SU(N)$ case were also conjectured by Jaewon Song.}
\begin{align} \label{okazakismith}
\mathbb{I}_{SU(N)}(q) = \frac{(q^{2N}, q^{2N})^N}{(q^2,q^2)} = \sum_{ \substack{\textbf{m} \in \mathbb{Z}^N \\ \textbf{m} \cdot \textbf{1} = 0 }} q^{N \textbf{m}^2 + 2 \mathbf{b} \cdot \mathbf{m}}
    \end{align}
    with $\mathbf{b} = (N{-}1,\cdots,1,0)$. Remarkably, it coincides with the generating functions of $N$-core partitions:
    \begin{align} \label{suNcore}
\mathbb{I}_{SU(N)}(q) = \sum_{k=0}^{\infty} c_N(k) q^{2k}\,,
    \end{align}
where the coefficient $c_N(k)$ is the number of integer partition of $k$ in terms of $n$-cores. Recall that an integer partition is an $N$-core if and only if it has no hook length divisible by $N$.  
    
Our goal here is to generalize this structure and establish exact formulas for all gauge groups $G$. The summation form in Eq.~\eqref{okazakismith} already suggests a Lie-theoretic pattern: the parameter $N$ may be identified with the Coxeter number $h$ (or the dual Coxeter number $h^\vee$), the vector $\mathbf{b}$ with the (dual) Weyl vector, and $\mathbf{m}$ with elements of the root or co-root lattice. For the simply-laced cases, the conjectured generalized formulas---in both product and summation forms---can be naturally related to
specializations of the Macdonald identities~\cite{macdonald72}. Moreover, the combinatorics of $N$-cores generalize to the enumeration of certain elements in affine Weyl groups~\cite{lecouveyW25}.  While the correctness of these conjectural formulas can be verified through
symbolic and numerical evaluation of the defining integration formulas, a conceptual derivation and general proof remain desirable. Furthermore, for \emph{non-simply-laced} Lie algebras, there are intrinsic ambiguities in formulating the appropriate conjectural expressions—closely related to the Langlands dual pairs of root systems. Both the product formula and the precise relation to combinatorial objects generalizing $N$-cores remain even more obscure.

Remarkably, we have succeeded in exactly evaluating the integration formulas by
applying the \emph{Macdonald identities}, thereby deriving exact closed-form
expressions for all simple gauge groups $G$ in a uniform manner. The affine
\emph{denominator} formula for the \emph{untwisted} affine Lie algebras plays a
central role in this analysis. The Macdonald identities assert the equality
between the infinite product and infinite sum representations of the affine
denominators. In the simplest case corresponding to $SU(2)$ or $A_1^{(1)}$, this
reduces precisely to the classical \emph{Jacobi triple product} identity.  In the integration formula of the superconformal indices, the integrands are
written in terms of the product forms of the affine denominators, and applying
the Macdonald identities allows term-wise evaluation, yielding results expressed
as $q$-series. More intriguingly, the results coincide with a particular
specialization of affine denominators; consequently, by the Macdonald
identities, they can also be expressed in product form as \emph{eta-quotients}
(products of Dedekind $\eta$-functions with integer exponents), which exhibit
interesting modular properties. Our results also establish a close connection
with the combinatorics of affine Grassmannian elements~\cite{lamLMSSZ14}, which
generalize the notion of $N$-core partitions to all affine Weyl groups.

Along the way, by employing similar techniques, we also derived exact formulas
for the evaluation of the integration formula defining the \emph{half} Schur
index in 4d $\mathcal{N}=2$ SYM theories. In addition to these
closed-form results, our analysis reveals a remarkable bilinear structure: the
full Schur index can be expressed as a bilinear pairing of half Schur indices
with insertions of Wilson lines. Schematically,
\begin{align}
  \mathbb{I}_{\mathrm{Schur}}
  \,=\,
  \sum_{\lambda}
  \mathrm{I}^{\lambda}_{\mathrm{half}}
  \, \mathrm{I}^{\lambda}_{\mathrm{half}} ,
\end{align}
where the sum runs over dominant integral weights $\lambda$ (equivalently, over
Wilson line insertions), and $\mathbb{I}_{\mathrm{half}}^{\lambda}$ denotes the
Neumann half Schur index decorated by a Wilson line labelled by the weight
$\lambda$. This bilinear structure can be understood more concretely in terms
of \emph{reproducing kernels} inserted into the index integrand. Such an
operation realizes the full Schur index as the gluing of two half indices and
admits multiple interpretations in different frameworks. In the following, we
will further develop and contextualize these results from several complementary
perspectives, including holomorphic twists and Poisson vertex algebras,
categorification via BPS defects, Schur quantization of the $K$-theoretical
Coulomb branch, IR formulas for BPS spectra, and class~$\mathcal{S}$
TQFT wave functions.

\paragraph{\underline{Holomorphic twist}}
The superconformal index is a powerful tool for probing the strong-coupling
dynamics and dualities of supersymmetric quantum field theories in four dimensions. It admits
two equivalent constructions: as the Witten index on $S^3 \times \mathbb{R}$
\cite{romelsberger05}, capturing the spectrum of states preserved by a
supercharge, and as the graded character of the cohomology classes of local
operators, arising from a chosen twisting supercharge in holomorphic or
holomorphic–topological twists (see,
e.g.,~\cite{kapustin06,eagerS18,saberiW19,budzikGKWWY23}). The latter approach, which
implements a cohomological reduction of the field theory rather than a full
supergravity background, applies for instance to $4d$ $\mathcal{N}=1$ theories
on $\mathbb{C}^2$ and to $\mathcal{N}=2$ theories on $\mathbb{C} \times
\mathbb{R}^2$. In particular, it is meaningful for both superconformal and
asymptotically free theories. 
In the asymptotically free case, however, the operator-counting interpretation
in twisted theories is generally restricted to indices graded by spins along the
holomorphic planes. For example, in $4d$ $\mathcal{N}=1$ theories, this
restriction allows for an index depending on two geometric parameters $(p,q)$,
whereas in $\mathcal{N}=2$ theories only the \emph{Schur index}—depending solely
on $q$—admits an operator-counting interpretation. Note that the full $\mathcal{N}=2$ superconformal index, defined as a Witten index depending on three fugacities $(p,q,t)$, can be defined even for non-conformal theories. However, its
interpretation as an operator-counting formula via the state–operator
correspondence in radial quantization is available only for SCFTs, where it
captures the contributions of short multiplets of the superconformal algebra.
Furthermore, SCFTs possess a $U(1)_r$ R-symmetry essential for supersymmetric
localization, since it permits a rigid off-shell supergravity background
compatible with the chosen supercharge \cite{festuccia11}. In contrast, for non-conformal theories no analogous localization framework is known. Therefore, in this work we adopt
the construction based on holomorphic and holomorphic–topological twists as a
robust framework for treating non-conformal cases. Similar constructions apply
also to the half-Schur index \cite{dimofteGG11}, either as a Witten index on $HS^3 \times
\mathbb{R}$ (where $HS^3$ denotes the hemisphere) or as a
holomorphic–topological twist on $\mathbb{C} \times \mathbb{R} \times
\mathbb{R}_{\ge 0}$, both subject to various boundary conditions.

In the $4d$ $\mathcal{N}=2$ case, the cohomology classes of local
operators carry a Poisson vertex algebra structure\footnote{It can be
further enhanced to a vertex operator algebra in the case of
$4d$ $\mathcal{N}=2$ SCFTs~\cite{beemLLPRvR13, OhY19, jeong19}.}, and the Schur
index of $4d$ $\mathcal{N}=2$ SYM can be identified with the
gauged partition function of complex chiral $bc$ ghosts~\cite{OhY19,niu21}.
The Neumann half-Schur index, in turn, corresponds to the gauged
partition function of real chiral $bc$ ghosts. To understand the bilinear
product structure, one may view the complex chiral $bc$ ghosts of $G$ as
decomposing into two real chiral $bc$ systems. The reproducing kernel then
plays the role of a Dirac delta function, identifying the gauge fugacities
of the two sets of ghosts.

\paragraph{\underline{Defects and categorification}}
Superconformal indices can be further extended by including insertions of
BPS defect operators, such as surface and line defects along the topological
$\mathbb{R}^2$ \cite{cordovaGS16}. These defect-enriched indices also admit an operator-counting
interpretation within the holomorphic–topological twist framework. The
holomorphic–topological twist for $\mathcal{N}=2$ theories assigns meaning to
superconformal indices with line or surface defects inserted in the
topological $\mathbb{R}^2$ as counting \emph{defect-changing} local operators.

This perspective naturally suggests a \emph{categorification} of the
superconformal index, treating defects of various dimensions in a unified way:
morphisms in the higher categories correspond to operators supported on
junctions of defects. In this picture, one obtains a 2-category whose objects
are surface defects, while the bulk lines form a category whose Grothendieck
algebra of simple objects reproduces the fusion algebra of BPS line operators.
The Schur index then appears as the trace over the endomorphisms of the
identity line. A geometric derivation using coherent sheaves on the affine
Grassmannian shows that the Schur index for $4d$ $\mathcal{N}=2$ SYM can be identified with the
partition function of gauged complex $bc$ ghosts~\cite{niu21}. Similarly, the Neumann
half-Schur index corresponds to the partition function of  gauged real $bc$
ghosts. Hence, the Poisson vertex algebra structure of the bulk local operators
emerges naturally from this categorical viewpoint.

Regarding the reproducing kernel underlying the bilinear structure, it may be
interpreted as a sheaf over the affine Grassmannian whose restriction to each
orbit corresponds to the bundle associated with
$\mathrm{End}(V_\lambda)\simeq V_\lambda\otimes V_\lambda^\vee$, for
$\lambda\in P_+$. In the defect language, this may be viewed as a domain wall
wrapping $\mathbb{C}$.

\paragraph{\underline{Schur quantization}}
The fusion algebra of BPS line operators is closely related to the Coulomb
branch of the $3d$ $\mathcal{N}=4$ KK theory obtained by compactifying the
$4d$ $\mathcal{N}=2$ theory on $\mathbb{R}^3 \times S^1$, known as the
$K$-theoretical Coulomb branch. The Schur quantization of this
$K$-theoretical Coulomb branch takes as input the fusion algebra of line
operators, together with a positive-definite inner product specified by the
Neumann half and full Schur indices in the presence of defects~\cite{gaiottoT24}.
This quantization produces a Hilbert space for a quantum mechanical system,
which turns out to be a quantum integrable system whose classical limit
coincides with the $K$-theoretic Coulomb branch as its phase space. For the
$4d$ $\mathcal{N}=2$ SYM theory with gauge group $G$, the corresponding
quantum integrable system is the relativistic open $G$ Toda system.
Furthermore, the Neumann half index corresponds to the \emph{real} Schur
quantization~\cite{gaiottoV24}, while the full Schur index corresponds to the
\emph{complex} Schur quantization. The splitting of the complex chiral $bc$
ghosts into a pair of real chiral $bc$ ghosts is consistent with this
structure and with the bilinear product.

\paragraph{\underline{IR formula}}
Since the superconformal index is a protected observable, it naturally probes the IR dynamics of the theory. In the case of the $4d$ $\mathcal{N}=2$ Schur index, this connection becomes especially transparent when viewed through the RG flow governed by the Seiberg–Witten solution on the Coulomb branch. A central feature of the IR physics on the Coulomb branch is the spectrum of BPS dyonic particles and their wall-crossing behavior. The cluster-algebraic structures of the BPS spectrum, captured by BPS quivers and cohomological Hall algebras~\cite{gaiottoGL24}, are related—via the Seiberg–Witten RG flow~\cite{ambrosinoG25}—to the fusion algebra of BPS line operators and its cluster algebra structure~\cite{gaiottoGL24}. An important wall-crossing object in this context is the Kontsevich–Soibelman (KS) monodromy operator, built from the BPS spectrum \cite{kontsevichS08}.

Remarkably, both the Schur index and the half-index turn out to be wall-crossing invariants, encoding essential information about the BPS spectrum. In particular, the rich algebraic structure of the BPS spectrum naturally induces corresponding IR expressions for the Schur and half-indices, which are closely related to the KS operator~\cite{cordovaGS16}. It is plausible that an analogous IR formula exists for the reproducing kernel itself, formulated in terms of quantum torus variables, thereby providing a natural explanation for the bilinear structure.

\paragraph{\underline{Class $\mathcal{S}$ and TQFT}} 
Both $4d$ $\mathcal{N}=1$ and $\mathcal{N}=2$ SYM theories admit a class~$\mathcal{S}$ realization via a partial topological twist of the $6d$ $(2,0)$ theory on a punctured Riemann surface with two identical irregular punctures~\cite{gaiottoMN09}. Through the duality chain of compactifications of the $(2,0)$ theory, the superconformal index of the resulting $4d$ theory can be expressed as the partition function of a two-dimensional TQFT defined on the punctured surface~\cite{gaddePRR09,gaddeRRY11}, which can further be identified with the partition function of a three-dimensional complex Chern–Simons theory~\cite{gaiottoV24}.

In this TQFT framework, each (regular or irregular) puncture is associated with a wave function labeled by a dominant weight, and the full index is obtained by gluing these wave functions. Remarkably, the wave functions for the irregular punctures relevant to SYM theories~\cite{song15} coincide with the Neumann half-Schur indices decorated by Wilson lines in the corresponding highest-weight representations, fully consistent with the correspondence implied by the real Schur quantization picture~\cite{gaiottoV24}.
Consequently, gluing two such wave functions reproduces the bilinear structure of the full Schur index, in precise agreement with our exact formulas. A similar bilinear structure also appears in the $4d$ superconformal index of $\mathcal{N}=1$ SYM theories, viewed as the mixed Schur index~\cite{song15}.

\paragraph{\underline{Organization of the paper:}} The paper is structured as follows. Section~\ref{section:mathpre} presents the relevant mathematical tools and results, focusing on Macdonald identities and their specializations. In Section~\ref{section:indices}, we introduce the integration formulas for the superconformal indices for 4d $\mathcal{N}=1$ and $\mathcal{N}=2$ SYM theories, carry out the explicit evaluation, and provide physical interpretations of the results.  Finally, Section~\ref{conclusion} summarizes our findings and discusses promising directions for future research.

\section{Macdonald Identities}\label{section:mathpre} 

In this section, we present several mathematical tools underlying the affine generalization of the Weyl denominator, which will later play a central role in the exact evaluation of superconformal indices. These include the Kac–Weyl character formula\footnote{For a physicist-friendly overview, see the appendix in \cite{gepnerW86}. Some useful mathematical monographs include \cite{kac3rd, wakimoto, carter_lie_2005}.}, the affine denominator formulas, and the Macdonald identities \cite{macdonald72}.  
We also discuss their specializations associated with inner automorphisms defined by regular semisimple elements \cite{kac78}, which lead to explicit eta-quotient expressions exhibiting modular properties—features related to the modular behavior of superconformal indices \cite{razamat12}.

For definiteness, we focus on the \emph{untwisted} affine Lie algebras $\mathfrak{g}^{(1)}$ in this work, although many statements extend naturally to the \emph{twisted} cases\footnote{The affine Lie algebras are classified as $X_r^{(t)}$ in Kac’s notation, where $X_r$ denotes a simple Lie algebra of type $X$ and rank $r$, and $t$ is the \emph{tier} number. When $t = 1$, $X_r^{(1)}$ corresponds to the \emph{untwisted} affine Lie algebras, which exist for all types $X_r$. The cases with $t \neq 1$ correspond to \emph{twisted} affine Lie algebras, which arise only for certain $X_r$ that admit nontrivial outer automorphisms.}. 
To fix notations, recall that the untwisted affine Lie algebra $\hat{\mathfrak{g}}^{(1)}$ can be realized as the central extension of the loop algebra
\begin{align}
    \hat{\mathfrak{g}}^{(1)} = \mathbb{C}K \oplus (\mathfrak{g} \otimes \mathbb{C}[z,z^{-1}]) \rtimes \mathbb{C}d, 
    \quad d = z\, \frac{d}{dz}\,,
\end{align}
where $\mathfrak{g}$ is the Lie algebra of a simple compact Lie group $G$, $K$ is the central element, and $d$ is the degree operator. The corresponding affine Cartan subalgebra is
$    \hat{\mathfrak{h}} = \mathfrak{h} \oplus \mathbb{C} K \oplus\mathbb{C}d$
with $\mathfrak{h}$ the Cartan subalgebra of $\mathfrak{g}$.

\subsection{Affine Lie algebra and Macdonald identity}

\noindent \textbf{Weyl character formula and denominator}  One of the most fundamental results in the theory of simple Lie algebras is the Weyl character formula:
\begin{align}
\chi_\lambda = \frac{J_{\lambda+\rho}}{J_\rho}\,, \quad J_{\lambda} = \sum_{w \in W} \varepsilon(w)\, w (e^{\lambda }).
\end{align}
and $\rho$ denotes half the sum of the positive roots, known as the Weyl vector, and $\varepsilon(w)$ is the signature of the Weyl group element $w$.
For a dominant integral weight $\lambda \in P_+$, $\chi_\lambda$ gives the character of the irreducible representation with highest weight $\lambda$. More generally, one has 
\begin{align}
     \chi_{w \cdot \lambda} = \varepsilon(w)\, \chi_\lambda,
\end{align}
where the shifted action of the Weyl group is defined by 
$w \cdot \lambda = w(\lambda + \rho) - \rho$.

These expressions may be regarded either as formal Laurent series or as genuine functions on a maximal torus $T$ of $G$. For a formal exponential $e^\alpha$ with $\alpha \in \mathfrak{h}^\ast$, the associated function on the maximal torus is given by
\begin{align}
    e^\alpha(s) = s^\alpha = \exp\!\big(2\pi i\,\langle \alpha,z \rangle\big),
\end{align}
where $\exp: \mathfrak{h}\to T$ sends $z \in \mathfrak{h}$ to $s=\exp(2\pi i z)$. Here we identify $\mathfrak{h}^\ast \simeq \mathfrak{h}$ by equipping $\mathfrak{h}^\ast$ with the standard inner product normalized such that $\langle \alpha,\alpha \rangle =2$ for all long roots $\alpha$. The Weyl group $W$ acts on the exponentials as
\begin{align}
    w(e^\alpha) = e^{w(\alpha)}\,, \quad w \in W.
\end{align}
It is also important to note that the Weyl numerator $J_{\lambda}$ is antisymmetric under the action of the Weyl group $W$,  
\begin{align}
    w(J_\lambda) = \varepsilon(w) J_\lambda \,, \quad w\in W\,.
\end{align}
At this stage, it is clear that the Weyl denominator $J_\rho$ is a special case of the numerator. On the other hand, $J_\rho$ also admits a product expression,
\begin{align}
J_\rho = e^{\rho }\prod_{\alpha \in R_+} (1 - e^{-\alpha}) = \prod_{\alpha \in R_+} (e^{\frac{\alpha}{2}} - e^{-\frac{\alpha}{2}})\,, 
\end{align}
where $R_+$ is the set of positive roots for $W$, giving rise to the Weyl denominator identity:
 \begin{align}
    \sum_{w \in W} \varepsilon(w)\, w (e^{ \rho}) = \prod_{\alpha \in R_+} (e^{\frac{\alpha}{2}} - e^{-\frac{\alpha}{2}})\,.
 \end{align}
Furthermore, the function \begin{align}
    \omega_H(s) := J_\rho(s) J_\rho(s^{-1}) = \prod_{\alpha \in R} (1 - s^{-\alpha})
\end{align} appears as the Haar measure in the Weyl integration formula for class functions on $G$:
\begin{align}
\int_G f(g)\, dg = \frac{1}{|W|} \oint_{T}\frac{ds}{(2\pi i)^r s}\, \omega_H(s)\, f(s)\,,
\end{align}
where $|W|$ is the order of the Weyl group, and $r = \mathrm{rank}(G)$. 
Here and throughout, we use the shorthand notation for the unit-circle
integration over the maximal torus $T$:
\begin{align}
  \oint_T \frac{ds}{(2\pi i)^r\, s}
  := \prod_{i=1}^r \oint_{|s_i|=1} \frac{ds_i}{2\pi i\, s_i},
\end{align}
where the contour of each $s_i$
encircles the unit circle once in the counterclockwise direction. Note that the Weyl integration formula underpins the integration formulae for superconformal indices in gauge-theoretic applications.

\noindent \textbf{Weyl-Kac character formula and affine denominator} Kac provides a generalization of the Weyl character formula to affine Lie algebras. 
In particular, for the dominant integral weight $\lambda$ of level $k$, i.e., $\lambda \in P_+$ such that $\langle \lambda, \theta \rangle \leq k$, there is an affine weight $\hat{\lambda} = \lambda + k \Lambda_0   \in \hat{\mathfrak{h}}^\ast$ where $\hat{\mathfrak{h}}^\ast = \mathfrak{h}^\ast \oplus \mathbb{R} \Lambda_0 \oplus \mathbb{R} \delta$ is the dual affine Cartan algebra for the $\mathfrak{g}^{(1)}$.
We denote by $\lambda = \overline{\hat{\lambda}}$ for the finite part in $\mathfrak{h}^\ast$ of $\hat{\lambda}$. 
The standard invariant bilinear form $\langle\cdot,\cdot \rangle$ on $\mathfrak{h}^\ast$ is extended to $\hat{\mathfrak{h}}^\ast$ by declaring
\begin{align}
\langle\Lambda_0,\delta\rangle=1, \qquad \langle\Lambda_0,\Lambda_0 \rangle=\langle\delta,\delta\rangle=\langle\Lambda_0,\alpha\rangle=\langle\delta,\alpha\rangle=0
\quad \text{for all } \alpha \in \mathfrak{h}^\ast. 
\end{align}
This extension makes $\langle\cdot,\cdot\rangle$ a non-degenerate symmetric bilinear form on $\hat{\mathfrak{h}}^\ast$, which in turn identifies $\hat{\mathfrak{h}}^\ast$ with $\hat{\mathfrak{h}}$. In particular, the elements $\delta$ and $\Lambda_0$ are identified with $d$ and $K$ respectively. 
Then the Weyl-Kac character takes the form
\begin{align}
\mathrm{ch}_{\hat{\lambda}}=\frac{J_{\hat{\lambda}+\hat{\rho}}}{J_{\hat{\rho}}}\,, \quad \text{with} \quad
J_{\hat{\lambda}}
= \sum_{w \in \hat{W}} \varepsilon(w) w( e^{\hat{\lambda} }) \,.
\end{align}
where $\hat{\rho} = \rho + h^\vee \Lambda_0$ and $\hat{W}$ is the affine Weyl group.
Note that the Kac numerator $J_{\hat{\lambda}}$ is antisymmetric under the action of $W \subset \hat{W}$:
\begin{align}
    w(J_{\hat \lambda}) = \varepsilon(w) J_{\hat{\lambda}} 
    \,, \quad  w\in W
    \,.
\end{align}
As mentioned before, these expressions can be treated either as formal series or functions on the maximal torus for the affine Lie group. In particular, for $\hat{\alpha} = \alpha + k \Lambda_0 - n \delta \in \hat{\mathfrak{h}}^\ast$
\begin{align}
    e^{\hat \alpha}(s,u,q) = u^k s^\alpha q^n\,.
\end{align}
Because the variable $u$ contributes only via multiplicative factors $u^k$, we specialize to $u=1$ to suppress this trivial dependence, retaining only the nontrivial dependence on $s$ and $q$. For brevity, we also write
\begin{align}
    \mathrm{ch}_{\hat{\lambda}}(s,q) := \mathrm{ch}_{\hat{\lambda}}(s,1,q)\,,
\end{align}
and adopt the same convention for other functions labelled by $\hat\lambda$.

On the one hand, the Kac denominator $  J_{\hat{\rho}}$ is a special case of the Kac numerator, on the other hand, the denominator $J_{\hat{\rho}}$ admits a product formula, 
\begin{align}
J_{\hat{\rho}}  = e^{\hat{\rho}}\prod_{\alpha \in \hat{R}_+} (1 - e^{-\alpha})^{\mathrm{mult}(\alpha)}
\end{align}
yielding the Weyl–Kac denominator identity:
\begin{align}
\sum_{w \in \hat{W}} \varepsilon(w) w( e^{\hat{\rho}}) = e^{\hat{\rho}}\prod_{\alpha \in \hat{R}_+} (1 - e^{-\alpha})^{\mathrm{mult}(\alpha)}
\end{align}
where $\hat{R}_+$ is the positive affine root system for $\hat{W}$, and $\mathrm{mult}(\alpha)$ is the multiplicity of an affine root $\alpha$. Note that the affine root system is given by
\begin{align} \label{eq:affine_root_system}
\hat{R} = \{ \alpha + k\delta \mid \alpha \in R,\, k \in \mathbb{Z} \} \cup \{ k\delta \mid k \in \mathbb{Z} \setminus \{0\} \}.
\end{align}
 Here each root of the form $\alpha + k\delta$ (with $\alpha \in R$) has multiplicity one, and each imaginary root $k\delta$ has multiplicity equal to the rank $r$ of the finite Lie algebra $\mathfrak{g}$.
The set of positive affine roots is
\begin{align} \label{eq:positive_affine_roots}
\hat{R}_+ = R_+ \cup \{ \alpha + k\delta \mid \alpha \in R,\, k > 0 \} \cup \{ k\delta \mid k > 0 \} \,.
\end{align}
 Furthermore, the ratio between the Kac denominator and the Weyl denominator, which we refer to as the \emph{Macdonald denominator}, is invariant under the action of the finite Weyl group  $W \subset \hat{W}$, is given by
\begin{align} \label{macJ}
    \mathcal{M} = \frac{J_{\hat{\rho}}}{J_\rho}  = e^{h^\vee \Lambda_0} \prod_{\alpha \in \hat{R}_+\backslash R_+} (1-e^{\alpha})^{\mathrm{mult}(\alpha)}\,, \quad w (\mathcal{M}) = \mathcal{M}  \quad \forall w \in W \,. 
\end{align}

Regarding the affine denominators as functions on the maximal torus of $G$, we further have the following explicit infinite product expressions
\begin{align}
J_{\hat{\rho}}(s,q) & = (q,q)^r \prod_{\alpha \in R_{+}} (s^\alpha q,q) (s^{-\alpha},q) = (q,q)^r \prod_{\alpha \in R_{+}} \theta (s^{-\alpha},q)\\
    \mathcal{M}(s,q) & = (q,q)^r \prod_{\alpha\in R} (s^\alpha q,q)\,.
\end{align}
Here $(x,q) = \prod_{k=0}^{\infty}(1-q^{k} x)$ is the $q$-Pochhammer symbol, $\theta(x,q) = (x,q)(x^{-1}q,q)$ is the theta function, and the factor $ (q,q)^r $ encodes the positive  \emph{imaginary} roots $\{n \delta \mid n>0\}$ each of which have a multiplicity of $r$. 

Furthermore, the infinite product formula for $\mathcal{M}(s,q)$ can also been identified with 
\begin{align}
    \mathcal{M}(s,q) = \prod_{i=1}^{\infty} \det{\!}_{\mathfrak{g}}\left( 1-q^n \mathrm{Ad}(s)\right)
\end{align}
which is defined in the adjoint representation of $\mathfrak{g}$ and $\mathrm{Ad}(s)$ is the adjoint action by the maximal torus element $s$. The following subsection makes essential use of this expression in the specialization of $\mathcal{M}(s,q)$.

\noindent \textbf{Macdonald identities}
As noted, the affine denominator—either $J_{\hat{\rho}}$ or $\mathcal{M}$—admits both summation and product expressions, whose equality is referred to as the \emph{Macdonald identities}.
The summation over the affine Weyl group $\hat W$ can be reorganized in two
natural ways.  First, using the semidirect product structure
$\hat W = Q^\vee \rtimes W$ with the coroot lattice $Q^\vee$ as a normal subgroup, every element
admits a unique decomposition $\hat w = \beta u$ with $\beta \in Q^\vee$ and
$u \in W$, so that
\begin{align}
       \sum_{\hat w \in \hat W} f(\hat w)
   = \sum_{u \in W} \left( \sum_{\beta \in Q^\vee} f(\beta u) \right).
\end{align}
Alternatively, viewing $W$ as a parabolic subgroup of $\hat W$, we have the
coset decomposition $\hat W = \bigsqcup_{\gamma \in \hat W/W} \gamma W$, and
hence
\begin{align}
   \sum_{\hat w \in \hat W} f(\hat w)
   = \sum_{\gamma \in \hat W/W} \left( \sum_{u \in W} f(\gamma u)\right).
\end{align}
In both descriptions, the sum over $W$ appears as an orbit sum in
$\mathcal M$, and as a signed sum in $J_{\hat \rho}$. Furthermore, in the first description, the outer sum is canonically over $W$, while in the
second case it depends on the choice of coset representatives of $\hat{W}/W$. 
Regarding the
choices of representatives for $\hat W/W$, the most obvious choice is given by the coroot lattice $Q^\vee$, which parametrizes the cosets in $\hat W/W$, each coset being represented uniquely by a translation element $t_\gamma$ with $\gamma \in Q^\vee$. Another choice we will consider is the set of \emph{affine Grassmannian elements}, denoted by $\hat W^0$, consisting of minimal length representatives in each coset $t_\gamma W$. Each $c \in \hat{W}^0$ admit decompostion $c = t_{\gamma} u =  u t_{\nu}$ where $u \in W$ is the unique element such that $\nu = u^{-1}(\gamma)$ is anti-dominant. This in turn implies that $\lambda = -h^\vee \nu + u^{-1}(\rho) - \rho = u^{-1}\cdot (-h^\vee \beta)$ is dominant \cite{lecouveyW25}. 
Altogether, these considerations lead to the following bijective descriptions of $\hat{W}/W$:
\begin{align}
    \hat{W}/W \simeq Q^\vee \simeq Q^\vee \cap (-P_+)\simeq \hat W^0 \simeq (W\cdot h^\vee Q^\vee)\cap P_+\,.
\end{align}

Below we summarize the different formulations of these identities and related aspects
\begin{itemize}
    \item 
The first involves $\hat{W}/W \simeq Q^\vee$, originally obtained by Macdonald in his celebrated work on affine root systems \cite{macdonald72}, which predates the development of affine Kac–Moody algebras:
\begin{align} \label{maciden1}
    \mathcal{M}(s,q) & = \sum_{\gamma  \in Q^\vee}  q^{\frac{1}{2}h^\vee \langle \gamma, \gamma \rangle + \langle \rho, \gamma \rangle} \chi_{h^\vee \gamma}(s)\\
    J_{\hat{\rho}} (s,q) & = \sum_{\gamma \in Q^\vee} q^{\frac{1}{2}h^\vee \langle \gamma, \gamma \rangle + \langle \rho, \gamma \rangle} J_{\rho+h^\vee \gamma}(s)
    \,.
\end{align}

\item The second is the Kostant-Fegan form, involving summation over the dominant weights \cite{kostant76, fegan78},
\begin{align} \label{maciden2}
    \mathcal{M}(s,q) & = \sum_{\lambda  \in P_{+}} \chi_{\lambda}(a) \chi_{\lambda}(s) q^{\frac{1}{2h^\vee}c_2(\lambda)} \\
    J_{\hat{\rho}}(s,q) & = \sum_{\lambda  \in P_{+}} \chi_{\lambda}(a) J_{\lambda + \rho}(s) q^{\frac{1}{2h^\vee}c_2(\lambda)} 
\end{align}
Here, $c_2(\lambda) = |\lambda + \rho|^2 - |\rho|^2$ is the quadratic Casimir value such that $c_2(\theta) = 2h^\vee$ for the highest root $\theta$, and the element $a = \exp\left(2\pi i \frac{\rho}{h^\vee}\right)$ is Kostant’s principal element of type $\rho$.

Note that $\chi_\lambda(a)$ takes values only in $\{0, \pm 1\}$, and is nonzero only when $\lambda$ is the unique dominant representative of $h^\vee \gamma$ for some $\gamma \in Q^\vee$ under the shifted Weyl action $u \cdot \gamma := u(\gamma + \rho) - \rho$, and $\chi_{\lambda}(a) = \varepsilon(u)$. Hence this involves $\hat{W}/W \simeq  (W\cdot h^\vee Q^\vee)\cap P_{+}$. Hence,
\begin{align} \label{maciden22}
    \mathcal{M}(s,q) & = \sum_{\lambda  \in (W\cdot h^\vee Q^\vee)\cap P_{+}} \varepsilon(u) \chi_{\lambda}(s) q^{\frac{1}{2h^\vee}c_2(\lambda)} \\
    J_{\hat{\rho}}(s,q) & = \sum_{\lambda  \in (W\cdot h^\vee Q^\vee)\cap P_{+}} \varepsilon(u)  J_{\lambda + \rho}(s) q^{\frac{1}{2h^\vee}c_2(\lambda)} 
\end{align}
with $u \in W$ uniquely determined by 
$u\cdot \lambda \in h^\vee Q^\vee$. These are the strong form of the Kostant-Fegan form. 

\item The third is the affine Grassmannian form \cite{lecouveyW25}, involving $\hat{W}/W \simeq \hat{W}^0$. 
Then we have the expressions of the affine denominators:
\begin{align} \label{maciden3}
J_{\hat{\rho}}(s,q) & = \sum_{c \in \hat W^0} \varepsilon(c) J_{\overline{c^{-1}(\hat \rho)}}(s)  q^{(\Lambda_0 - c(\Lambda_0), \,\hat{\rho})}\\
\mathcal{M}(s,q) & = \sum_{c \in \hat W^0} \varepsilon(c) \chi_{\overline{c^{-1}(\hat \rho) -\hat \rho}}(s)  q^{(\Lambda_0 - c(\Lambda_0), \,\hat{\rho})}
\end{align}
Note that for $c = t_\gamma u = u t_\nu $ where $\nu = u^{-1}(\gamma)$ is an anti-dominant weight such that $c$ is the minimal length representative in $\gamma W$, we have 
\begin{align}
        \Lambda_0 - c(\Lambda_0) & = \frac{1}{2} |\gamma|^2 \delta - \gamma\\
       ( \Lambda_0 - c(\Lambda_0) , \hat{\rho}) & = \frac{h^\vee}{2} |\gamma|^2 - \langle\gamma,\rho \rangle
    \end{align}
On the other hand, $c^{-1} = t_{-\nu} u^{-1}$,
\begin{align}
    c^{-1}(\hat{\rho}) - \hat{\rho} = - h^\vee \nu + u^{-1}(\rho) - \rho - \left( \frac{h^\vee}{2} |\nu|^2 - \langle u(\nu), \rho\rangle \right)\delta\,. 
\end{align}
Hence,
\begin{align}
    \overline{c^{-1}(\hat{\rho}) - \hat{\rho}} & = - h^\vee \nu + u^{-1}(\rho) - \rho = u^{-1} \cdot (-h^\vee \gamma)
\end{align}
is a dominant weight $\lambda \in P_+$. Furthermore, 
\begin{align}
\langle  \hat{\rho} - c^{-1}(\hat{\rho}), \Lambda_0 \rangle  & = \frac{h^\vee}{2} |\gamma|^2 - \langle \gamma, \rho\rangle = \frac{1}{2 h^\vee} c_2(\lambda)\,.
\end{align}
We notice that
\begin{align}
   \langle  \hat{\rho} - c^{-1}(\hat{\rho}), \Lambda_0 \rangle )   = ( \Lambda_0 - c(\Lambda_0) , \hat{\rho}) \,.
\end{align}
Following these calculations, replacing $\gamma$ by $-\gamma$ makes the equivalence with the strong Kostant–Fegan form \eqref{maciden22} and with the lattice–sum representation \eqref{maciden1} explicit.

\item The fourth is the classical theta function form, involving the semi-direct product decomposition of $\hat{W}$.  More generally, for $\hat{\lambda} = \lambda + k \Lambda_0$, the Kac numerator is given by
\begin{align}
    J_{\hat \lambda}(s,q) & = q^{\frac{|\lambda+\rho|^2}{2(k+ h^\vee)}} \sum_{w \in W} \varepsilon(w) \Theta_{w(\hat{\lambda})}(s,q)\,,
\end{align}
with the classical theta function defined by
\begin{align}
    \Theta_{\hat{\lambda}}(s,q) = \sum_{\gamma \in Q^\vee } q^{\frac{k}{2} |\lambda +k \gamma|^2} s^{\lambda+k\gamma}\,.
\end{align}
Hence the affine denominators are written as
\begin{align}
    J_{\hat \rho}(s,q) & = q^{\frac{|\rho|^2}{2 h^\vee}} \sum_{w \in W}  \varepsilon(w) \Theta_{w(\hat{\rho})}(s,q) \\
    \mathcal{M}(s,q) & = q^{\frac{|\rho|^2}{2 h^\vee }} \sum_{w \in W} \frac{\Theta_{w(\hat{\lambda})}(s,q)}{J_{w(\rho)}(s,q)}
\end{align}
Note that the appearance of classical theta functions indicates the underlying modularity.

\end{itemize}

Below we make some further comments:
\begin{itemize}
    \item Note that $\mathcal{M}(s,q)$ is related to the heat kernel on compact Lie groups. It can be identified with a specialization of Frenkel's formula for the  \emph{central} heat kernel \cite{frenkel84}
    \begin{align}
\mathcal{F}(t,s,q)=\sum_{\lambda \in P_+} \chi_\lambda(t) \chi_\lambda(s) q^{\frac{1}{2 h^\vee}c_2(\lambda)}\,,
\end{align}
hence giving to the Kostant-Fegan form
\begin{align}
         \mathcal{M}(s,q):= \mathcal{F}(a,s,q) \,.
    \end{align}
where $a$ is Kostant’s principal element of type $\rho$.
Frenkel’s central heat kernel is known to admit physical derivations either from the partition function of 1d SUSY sigma-model on a group manifold $G$ \cite{choiT25}, or from the partition function of 2d Yang–Mills theory on a cylinder with simple gauge group $G$, as given by Migdal’s formula \cite{witten92}.

\item One interesting aspect of the affine Grassmannian formulation is its connection to the combinatorics of affine Weyl groups, generalizing the notion of $n$-core partition functions in the type $A$ case \cite{garvanKS90}. Concretely, an affine Grassmannian element $c$ corresponds to an $n$-core partition, while the coroot lattice vector $\gamma \in Q^\vee$ parametrizes the data in terms of Maya diagrams and multicharges in a fermionic description. Such relations have been established for the seven infinite series of affine Lie algebras of classical type\footnote{For a study of Macdonald identities for these affine Lie algebras 
using elliptic determinant evaluations, see \cite{rosengrenS05}. It is also closely related to the combinatorial approach of \cite{stanton89}. }, as well as for the affine $G_2$ case, encompassing both \emph{untwisted} and \emph{twisted} types.

The Macdonald denominator $\mathcal{M}(s,q)$, viewed as a $q$-series, can be identified with the multi-parameter generating function of affine Grassmannian elements. Depending on the chosen bijective description, one obtains two equivalent realizations: either as an infinite product $\prod$ or as an infinite vector sum $\sum$.
\begin{figure}[H]
    \centering
\begin{tikzpicture}[every node/.style={draw, rectangle, rounded corners, minimum width=1.5cm, minimum height=0.8cm, align=center, fill=gray!10}]

  \node (A) at (0,0) {$\prod$};
  \node (B) at (3,0) {$\sum$};
  \node (C) at ($(A)!0.5!(B)+(0,2)$) {$\mathcal{M}$};

  \draw (A) -- (B) -- (C) -- (A);

  \path (C) -- (A) node[midway, draw, circle, fill=white, minimum size=15pt, inner sep=0pt] {$3$};
  \path (A) -- (B) node[midway, draw, circle, fill=white, minimum size=15pt, inner sep=0pt] {$1$};
  \path (B) -- (C) node[midway, draw, circle, fill=white, minimum size=15pt, inner sep=0pt] {$2$};
\end{tikzpicture}
    \caption{The relations between generalized $n$-cores partition functions and Macdonald identity.}
\label{fig:core}
\label{fig:core}
\end{figure}
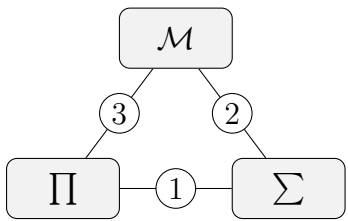
As illustrated in Fig.\ref{fig:core}, edge $1$ represents the Macdonald identities, edge~$2$ encodes the bijection via Littlewood decomposition for certain classes of partitions, and edge~$3$ corresponds to the two bijective descriptions of affine Grassmannian elements.

Another noteworthy aspect is that affine Grassmannian elements in $\hat{W}^0$ (and more generally $\hat{W}/W$) parametrize the Schubert cell decomposition of the affine Grassmannian $\mathrm{Gr}(\hat{G})$ which can be defined as
\begin{align}
    \mathrm{Gr}(\hat{G}) = G((z))/G[[z]]
\end{align}
where $G((z))$ denotes the loop group of $G$ over formal Laurent series, and 
$G[[z]]$ denotes the subgroup over formal power series. 
Such a connection suggests a link between our results involving summation over the coset $\hat{W}/W$ and the combinatorics and geometry of $\mathrm{Gr}(\hat{G})$ involving affine Schubert calculus \cite{lamLMSSZ14}.

\end{itemize}

\subsection{Specialization of Macdonald identities}
Generically, the Macdonald identities involve several parameters associated with the maximal torus of $G$. Specializing these parameters yields interesting $q$-series identities expressed in terms of Dedekind eta-quotients \cite{macdonald72, kac78}. Particularly notable cases arise when the specialization corresponds to finite-order regular semisimple elements, which induce inner automorphisms of $\mathfrak{g}$ via the adjoint action. For any such an element $\sigma$, one further requires that its characteristic polynomial $\det_{\mathfrak{g}}(1 - q\,\mathrm{Ad}(\sigma))$ have integer coefficients, a condition that characterizes \emph{quasi-rational} automorphisms \cite{kac78}. 
For a quasi-rational inner automorphism $\sigma$ of order $k$, the Lie algebra $\mathfrak{g}$ decomposes as
\begin{align}
    \mathfrak{g} = \bigoplus_{i=0}^{k-1} \mathfrak{g}_i \,,
\end{align}
where $\mathfrak{g}_i$ is the eigenspace of $\sigma$ with eigenvalue $\omega^i$ where $\omega = e^{2\pi i/k}$. In particular, $\mathfrak{g}_0$ is identified with $\mathfrak{h}$. 
Let $d_i = \dim(\mathfrak{g}_i)$, $i = 0, 1,\cdots, k-1$.  Hence, $d_0 = r$, $\dim G = \sum_{i=0}^{k-1} d_i$. Then the characteristic polynomial can be written as
\begin{align}
    \det {\!}_{\mathfrak{g}}(1-q \sigma) = (1-q)^r \prod_{i=1}^{k-1} (1-\omega^{d_i} q) = \prod_{i \mid k} (1-q^i )^{n_{k/i}}\,,
\end{align}
where the set of integers $\{n_{i} \}$ are related to the $\{d_i\}$ via M\"obius reversion
\begin{align}
    n_i = \sum_{j \mid i} d_j \mu(\frac{i}{j}). 
\end{align}
where $\mu(n)$ is the M\"obius function
\begin{align}
\mu(n) =
\begin{cases}
  1 & \text{if } n = 1, \\[6pt]
  (-1)^k & \text{if } n \text{ is the product of $k$ distinct primes}, \\[6pt]
  0 & \text{if } n \text{ is divisible by a square greater than } 1.
\end{cases}
\end{align}
Conversely, $d_i = \sum_{j \mid i} n_j$. 

Recall that the Macdonald formula can be expressed as
\begin{align*}
    \mathcal{M}(z,q) = 
    \prod_{n=1}^{\infty} 
    \det{\!}_{\mathfrak{g}}(1-q^{n}\mathrm{Ad}(z))\,.
\end{align*}
Then for a group element $\sigma$ corresponding to a quasirational automorphism of order $k$, the specialization of the Macdonald formula at $\sigma$ admits the product expression 
\begin{align}
    \mathcal{M}(\sigma,q) = \prod_{i \mid k} (q^i, q^i )^{n_{k/i}}\,.
\end{align}

Further in terms of Dedkind eta-function $\eta(q) = q^{\frac{1}{24}} (q,q)$, we can get
\begin{align}
   \mathcal{M}(\sigma,q) = q^{-\frac{1}{24}\dim G} \prod_{i\mid k} \eta(q^{i})^{n_{k/i}}. 
\end{align}
where we have used the relation
\begin{align}
   \sum_{i|k} i \,n_{k/i} = \dim G \,. 
\end{align}
So the results are expressed in terms of eta-quotients with conductor $k$. 

Now we apply these to the case of $\sigma = \mathrm{Ad}(a)$, which has order $k = l h^\vee$ with $l$ the lacing number. In the table below, we list some data associated to $G$:
\begin{table}[h]
\centering
\begin{tabular}{|c|c|c|c|c|c|c|c|c|c|}
\hline
$G$ & $A_n$ & $B_n$ & $C_n$ & $D_n$ & $E_6$ & $E_7$ & $E_8$ & $F_4$ & $G_2$ \\
\hline
$\dim$ & $n^2+2n$ & $2n^2+n$ & $2n^2+n$ & $2n^2-n$ & $78$ & $133$ & $248$ & $52$ & $14$ \\
\hline
$h^\vee$ & $n+1$  & $2n-1$  & $n+1$ & $2n-2$ & 12 & 18 & 30 & 9 & 4 \\
\hline
$l$ & 1 & 2 & 2 & 1 & 1 &  1 & 1 & 2 & 3 \\
\hline
\end{tabular}
\caption{Data of simple Lie groups $G$. }
\end{table}

First, we have the following result for $\{d_i\}$ and $\{n_i\}$:
\begin{itemize}
    \item{simply-laced cases:} We have $l=1$, and $h^\vee = h$ the Coxeter number.  
    For $\alpha \in R_+$, the $\langle \alpha, \rho \rangle$ is the height of $\alpha$. 
    The partition of the multiplicities of heights of positive roots $\{r \equiv \eta_1 \geq \eta_2 \cdots \geq  \eta_{h-1} \equiv 1\}$ is conjugate to the partition of the exponents $\{1 \equiv m_1 \leq m_2 \leq \cdots \leq m_r \equiv h-1\}$. Since the exponents satisfies $m_i + m_{j} = h$ if $i+j = r+1$, it leads to $\eta_p + \eta_q = r$ if $p+q = h+1$.  
    Further note that $\eta_p = \eta_{p+1} + \chi_{\{m_i\}}(p) $ where $\chi_{\{m_i\}} (p)$ is 1 if $p \in m_i$, is 0 othewise.  Extending to $R$, we also have $\eta_{-i} = \eta_i$ with $i =1,2,\cdots, h-1$. Then \cite{macdonald72}
    \begin{align}
        d_p = \eta_p + \eta_{-h+p}= \eta_p + \eta_{h-p} = \eta_{p+1} + \eta_{h-p} + \chi_{\{m_p\}}(i) = l + \chi_{\{m_p\}}(p)\,,
    \end{align}
where $ p =0, 1, \cdots, h-1 $.
Then the character polynomials can be written as
    \begin{align}
        (1-q^h)^r \prod_{i=1}^{r} (1-\omega^{m_i}q)\,.
    \end{align}
    
\item{non simply-laced cases:} 
By explicit computation, we find that \begin{itemize}
    \item For $B_n$, $l h^\vee = 4n-2$, we find
\begin{align}
          d_{0} & = n, \quad d_{2n-1} = 2,\\
      d_i & = d_{4n-2-i} = 
      \begin{cases}
        1 & i \text{ odd}, \\[2pt]
        n & i \text{ even},
      \end{cases}
      \quad 1 \leq i < 2n-1.
\end{align}
Furthermore, 
\begin{align}
    n_1 =1, \quad n_2 = n-1, \quad n_{2n-1} =1, \quad n_{4n-2} = -1
\end{align}

\item For $C_n$, $l h^\vee = 2n+2$, we find 
\begin{align}
   d_{0} & = n, \quad 
      d_{n+1} = 2n - 2\!\left\lfloor \tfrac{n}{2} \right\rfloor, \\
       d_i & = d_{2n+2-i}  = \begin{cases}
        n-1 & i \,\, \text{odd}\\
        n & i \,\, \text{even}\\
    \end{cases},  \quad \text{for} \quad 1\leq i < n+1 \,. 
\end{align}
Furthermore
\begin{align}
    n_1 =n-1, \quad n_2 = 1, \quad n_{n+1} =1, \quad n_{2n+2} = -1
\end{align}

\item For $F_4$,
\begin{align}
    \{d_i\mid i =0,1, \cdots, 17\} = \{4,2,4,1,4,2,4,2,4,2,4,2,4,2,4,1,4,2\}.
\end{align}
Furthermore,
\begin{align}
    n_1 =2, \quad n_2 = 2, \quad n_{3} =-1, \quad n_{6} = 1 , \quad n_{9} =1, \quad n_{18} = -1
\end{align}

\item For $G_2$,
\begin{align}
    \{d_i\mid i =0,1, \cdots, 11\} = \{2,1,0,2,1,1,2,1,1,2,0,1\}.
\end{align}
Furthermore,
\begin{align}
    n_1 =1, \quad n_2 = -1, \quad n_{3} =1, \quad n_{4} = 1 , \quad n_{6} =1, \quad n_{12} = -1
\end{align}
    
\end{itemize}

\end{itemize}

In terms of the results above, we further obtain the following uniform expressions for $\mathcal{M}$ for all $G$. In particular, the simply laced $ADE$ cases exhibit further uniformed pattern. 
\begin{itemize}
    \item For the simply-laced case,
\begin{align} \label{specADEMG}
\mathcal{M}(a,q) = \frac{(q^{h}, q^{h})^{r+1}}{(q, q )} f_G(q)
\end{align}
with 
\begin{align}
\label{specADEfG}
    f_G(q) = 
\begin{cases}
    1 & A_{n-1} , \, h= n\\
    \frac{(q^2, q^2)}{(q^{n-1},q^{n-1})}
    & D_{n} ,\, h = 2n-2\\
    \frac{(q^{3},q^{3})(q^{2},q^{2})}{(q^{6},q^{6})(q^{4},q^{4})}
    & E_{6} ,\, h = 12\\
    \frac{(q^{3},q^{3})(q^{2},q^{2})}{(q^{9},q^{9})(q^{6},q^{6})}
    & E_{7}, h =18 \\
    \frac{(q^{5},q^{5})(q^{3},q^{3})(q^{2},q^{2})}{(q^{15},q^{15})(q^{10},q^{10})(q^{6},q^{6})}
    & E_{8}, \, h =30 \\
\end{cases}
\end{align}

\item 
For the non-simply laced cases, 
\begin{align}
\label{specBCFGMG}   
\mathcal{M}(a,q) = (q^{l h^{\vee}},q^{l h^{\vee}})^{n_s} (q^{ h^{\vee}},q^{ h^{\vee}})^{n_l} f_{G}(q),
\end{align}
with
\begin{align}
\label{specBCFGfG}
f_G(q) = 
\begin{cases}
   \dfrac{(q^2,q^2)}{(q,q)}  & B_{n}, C_{n}, \\
   \dfrac{(q^3, q^3)(q^2,q^2)}{(q^6,q^6)(q,q)} & F_4, G_2,
\end{cases}
\end{align}
where $n_s$ and $n_l$ denote the numbers of short and long simple roots, 
given respectively by $(1,n-1)$, $(n-1,1)$, $(2,2)$, and $(1,1)$ 
for $B_n$, $C_n$, $F_4$, and $G_2$.

\end{itemize}
Finally, following Macdonald's identity, these specialized infinite product formulas yield various summation formulas.

\section{SUSY Index for 4d Super Yang-Mill Theories}\label{section:indices}
In this section, we first define the integration formulas for the 4d $\mathcal{N}=1$ and $\mathcal{N}=2$ SYM theories. We then perform their exact evaluations using Macdonald identities, and finally discuss various interpretations of the results.

\subsection{Definition}
Our starting point is the working definition of the superconformal index given by its integral formulas. As mentioned in the introduction, it can be defined in terms of the Witten index using the Lagrangian description, even though a rigorous justification of the operator-counting meaning of this definition requires the powerful frameworks of holomorphic twists and holomorphic–topological twists. Note that a good review of the general aspects of the 4d superconformal index can be found in \cite{rastelliR14}, where the prescription for the integration formulas can be found.  

Now we present results for 4d $\mathcal{N}=1$ and $\mathcal{N}=2$ SYM theories. Firstly, in the $\mathcal{N}=1$ case, we consider the full index
\begin{align}
    I_G(p,q) = \frac{1}{|W|}\oint_{T} \frac{dz}{(2\pi i)^r z}\frac{(p,q)^r(q,q)^r }{\prod_{\alpha \in R} \Gamma(z^\alpha,p,q)}
\end{align}
where the elliptic Gamma function is defined by
\begin{align}
    \Gamma(z,p,q) = \mathrm{PE}[\frac{z - \frac{pq}{z}}{(1-p)(1-q)}]=\prod_{m,n=0}^{\infty}\frac{(1- p^{m+1} q^{n+1}/z)}{(1- p^m q^n z)}
\end{align}
With the identity
\begin{align}
    \Gamma(x,p,q)\Gamma(x^{-1},p,q) = \frac{1}{\theta(x,p)\theta(x^{-1},q)}\,,
\end{align}
the index can then be transformed into an expression with theta functions
\begin{align}
    I_G(p,q) = & \, \frac{1}{|W|}\oint_{T} \frac{dz}{(2\pi i)^r z} (p,q)^r(q,q)^r \prod_{\alpha \in R_+} \theta(z^\alpha,p)\theta(z^{-\alpha},q)\\
   = & \frac{1}{|W|}\oint_{T} \frac{dz}{(2\pi i)^r z} w_H(z)  (p,q)^r(q,q)^r \prod_{\alpha \in R} (p z^\alpha,p)(q z^{\alpha},q) 
\end{align}

Then, in the 4d $\mathcal{N}=2$ case, we consider the Schur index given by
\begin{align}
    \mathbb{I}_G(z, q) =  \frac{1}{|W|}\oint_{T} \frac{dz}{(2\pi i)^r z} w_H(z) \, \mathcalII_{\text{vec}}(z,q) \, ,
\end{align}
as well as the half index with Neumann boundary conditions with the insertion of Wilson line operators in $G$-module $R$ specified by
\begin{align}
    \mathrm{I}_G^R(z, q) =  \frac{1}{|W|}\oint_{T} \frac{dz}{(2\pi i)^r z} w_H(z) \, \mathcal{I}_{\text{vec}}(z,q) \chi_R(z)\, .
\end{align}
Here we have used the indices for the free vector multiplets
\begin{align}
    \mathcalII_{\text{vec}}(z,q) &\,= (q,q)^{2r} \prod_{\alpha \in R} (q z^{\alpha},q)^2\\
    \mathcal{I}_{\text{vec}} (z,q)&\,= (q,q)^r \prod_{\alpha \in R} (q z^{\alpha},q)
\end{align}
and these satisfy 
\begin{align}
    \mathcal{I}_{\text{vec}} (z,q) & =\, \mathcal{I}_{\text{vec}} (z^{-1},q)\\
    \mathcalII_{\text{vec}}(z,q) & =\, \mathcal{I}_{\text{vec}} (z,q)^2 = \mathcal{I}_{\text{vec}} (z,q) \mathcal{I}_{\text{vec}} (z^{-1},q)
\end{align}

We also note that the index with the Dirichlet boundary condition (with vanishing magnetic flux) can be identified with the Macdonald-Kac denominator $\mathcal{M}_G$.

We remark that the integration formulas studied here are closely related to, and in fact generalize, the well-studied elliptic beta integrals associated with root systems, which admit exact evaluations and have found a range of applications in physics \cite{spiridonov08}. In particular, the half Schur index can be obtained as a suitable limit of the elliptic beta integral defining the half Schur index for 4d $\mathcal{N}=4$ SYM theories which admits the integrand given by \cite{spiridonovV10}
\begin{align}
   (q,q)^r \prod_{\alpha \in R}\frac{(s^\alpha,q)}{(t s^\alpha,q)}\,.
\end{align}
Then $t \to 0$ limit produces the integrand for $\mathcal{N}=2$ SYMs
\begin{align}
    w_H(s)  \mathcal{M}_G(s,q)\,. 
\end{align}
We remark that this corresponds to the $q$-Whittaker limit of the Macdonald weight functions~\cite{noumi23}. 

The full Schur index, on the other hand, provides a genuine generalization of such elliptic beta integrals, whose exact evaluations remain largely unknown, except for the $SU(N)$ theory and a few lower-rank cases \cite{panP21, hatsudaO22}. In this work, however, we make progress toward evaluating these integrals for 4d $\mathcal{N}=1$ and  $\mathcal{N}=2$ super Yang–Mills theories with arbitrary simple gauge groups.


\subsection{Evaluation}
A key observation is the close connection between the integrands and the affine denominator formulas. In particular, using the $W$-symmetric Macdonald denominator $\mathcal{M}G$, the integration formulas can be rewritten as
\begin{align}
    I_G(p,q) & =\, \frac{1}{|W_G|} \oint_{T(G)} \frac{\mathrm{d}s}{(2 \pi i)^r s} \, w_H(s) \mathcal{M}_G(s, p) \mathcal{M}_G(s^{-1}, q) \\
    \mathbb{I}_G(q) & = \, \frac{1}{|W_G|} \oint_{T(G)} \frac{\mathrm{d}s}{(2 \pi i)^r s} \, w_H(s) \mathcal{M}_G(s, q)^2 \\
    \mathrm{I}_G^\lambda(q) & = \, \frac{1}{|W_G|} \oint_{T(G)} \frac{\mathrm{d}s}{(2 \pi i)^r s} \, w_H(s) \mathcal{M}_G(s, q) \chi_\lambda(s^{-1})
\end{align}
Similarly, using the $W$-skew-symmetric denominator $J_{\hat{\rho}}$, the same integrals take the form
\begin{align}
    I_G(p,q) & =\,  \frac{1}{|W_G|} \oint_{T(G)} \frac{\mathrm{d}s}{(2 \pi i)^r s} \,  J_{\hat{\rho}} (s, p) J_{\hat{\rho}} (s^{-1}, q)\\
    \mathbb{I}_G(q) & = \,  \frac{1}{|W_G|} \oint_{T(G)} \frac{\mathrm{d}s}{(2 \pi i)^r s} \,  J_{\hat{\rho}} (s, q) J_{\hat{\rho}} (s^{-1}, q)\\
   \mathrm{I}_G^\lambda(q) & = \,  \frac{1}{|W_G|} \oint_{T(G)} \frac{\mathrm{d}s}{(2 \pi i)^r s} \,  J_{\hat{\rho}} (s, p) J_{\lambda+\rho } (s^{-1}, q)
\end{align}
These two sets of expressions are related by the identity
\begin{align}
   J_{\hat{\rho}}(s,q) J_{\hat{\rho}}(s^{-1},q) = w_H(s) \mathcal{\mathcal{M}}(s,q)^2 \,.
\end{align}
which shows that the $W$-symmetric and $W$-skew-symmetric formulations are equivalent.

Now we can apply the various forms of the Macdonald identity to evaluate the integration. Let us first consider the Kostant-Fegan form \eqref{maciden2} for $\mathcal{M}_G$. Note that the characters $\chi_\gamma(z)$, indexed by dominant weights $\gamma \in P_+$, form an orthonormal basis with respect to the Haar measure on the maximal torus:
\begin{align}
    \frac{1}{|W_G|} \oint_{T(G)} \frac{\mathrm{d}s}{(2 \pi i)^r s} w_H(z)\, \chi_\lambda(z)\, \chi_{\lambda'}(z^{-1}) = \delta_{\lambda, \lambda'}\,, \quad \lambda, \lambda' \in P_+\,.
\end{align}
Hence, applying this to the integration formulas immediately leads to
\begin{align}
    I_G(p,q) &=\, \sum_{\lambda \in P_+} \chi_\lambda(a)^2\, (pq)^{\frac{1}{2h^\vee} c_2(\lambda)} \\
    \mathbb{I}_G(q) &= \sum_{\lambda \in P_+} \chi_\lambda(a)^2\, q^{\frac{1}{h^\vee} c_2(\lambda)} \\
    \mathrm{I}_G^\lambda(q) & = \, \chi_\lambda(a)q^{\frac{c_2(\lambda)}{2 h^\vee}} \label{defecthalfindex}
\end{align}
where $a$ is Kostant's principal element. We note that due to the presence of $\chi_\lambda(a)$, the summation can be restricted to the dominant image of the lattice $h^\vee Q^\vee$ under the shifted Weyl group action, namely $\left(W\cdot h^\vee Q^\vee \right)\cap P_{+}$, involving the strong
Kostant–Fegan form~Eq.~\eqref{maciden22}.

Remarkably, in the second equality of the first two rows, we can recognize the result as a specialization of the Macdonald denominator $\mathcal{M}$ at $s = a$, up to a redefinition of $q$ by $pq$ and $q^2$, respectively:
\begin{align}
I_G(p,q) &= \mathcal{M}_G(a, pq),\\
\mathbb{I}_G(q) &= \mathcal{M}_G(a, q^2).
\end{align}
Using Eqs.~\eqref{specADEMG}, \eqref{specADEfG}, \eqref{specBCFGMG}, and \eqref{specBCFGfG}, the corresponding product formulas follow immediately.
In particular, the case $G = SU(n)$ reproduces the product formula given in Eq.~\eqref{okazakismith}.

Since Kostant’s element $a$ satisfies $\chi_{h^\vee \gamma}(a) = 1$ for $\gamma \in Q^\vee$, then the specialization of Eq.~\eqref{maciden1} at $s=a$ leads to the coroot lattice summation formula
\begin{align} \label{latticeform}
        I_G(p,q) &= \sum_{\gamma \in Q^\vee} (pq)^{\frac{h^\vee}{2} \langle \gamma, \gamma \rangle + \langle \rho, \gamma \rangle} \\
        \mathbb{I}_G(q) &= \sum_{\gamma \in Q^\vee} q^{h^\vee \langle \gamma, \gamma \rangle + 2\langle \rho, \gamma \rangle}
\end{align}
Now let $\gamma = \sum_{i=1}^{r} m_i \alpha_i^\vee$, we can express the grading of $q$ as
\begin{align} \label{casimir}
\begin{split}
 \frac{h^\vee}{2} \langle \gamma, \gamma \rangle + \langle \rho, \gamma \rangle 
&= \frac{h^\vee}{2}  \sum_{i,j=1}^{r} m_i m_j G_{ij} + \sum_{i=1}^{r} m_i,
\end{split}
\end{align}
with
\begin{align}
    G_{ij} =\langle \alpha_i^\vee, \alpha_j^\vee \rangle= 2 A_{ij}/\langle \alpha_i, \alpha_i \rangle = \frac{4 \langle \alpha_i, \alpha_j \rangle}{\langle \alpha_i, \alpha_i \rangle \langle \alpha_j, \alpha_j \rangle}\,.
\end{align} 
Here, the matrix $(G_{ij})$ is given by a symmetrized form of the Cartan matrix $(A_{ij})$. 
   Hence, we can get a more explicit expression
\begin{align} \label{idxseries}
I_G(p,q) & = \sum_{m_1, \dots, m_r \in \mathbb{Z}} (pq)^{ \frac{h^\vee}{2} \sum_{i,j=1}^{r} m_i m_j G^{ij} +  \sum_{i=1}^{r} m_i}\, \\
\mathbb{I}_G(q) &= \sum_{m_1, \dots, m_r \in \mathbb{Z}} q^{ h^\vee \sum_{i,j=1}^{r} m_i m_j G^{ij} + 2 \sum_{i=1}^{r} m_i}\,.
\end{align}
In the special case $G = SU(n)$, the summation formula given in Eq.\eqref{okazakismith} is exactly reproduced.

Regarding the exact result for the half Schur index~\eqref{defecthalfindex}, we first note that for $\lambda = 0$, corresponding to the case without line insertions, the result is
\begin{align}
\mathrm{I}_G(q) \equiv \mathrm{I}_G^0(q) = 1\,,
\end{align}
which implies that only the identity operator contributes in the presence of the Neumann boundary condition.

Moreover, for any $k \in \mathbb{C}$ and any $\lambda,\mu \in \mathfrak{h}^\ast$, the Weyl numerator satisfies
\begin{align}
J_{\lambda}(e^{k\mu}) = J_{\mu}(e^{k\lambda})\,.
\end{align}
Then by Weyl character and denominator formulas, the following product formula holds for $a$, the Kostant's principle element,
\begin{align}
    \chi_{\lambda}(a) = \prod_{\alpha \in R_+} \frac{\sin\left(\frac{\pi}{h^\vee} \langle \lambda+ \rho,\alpha\rangle\right)}{\sin\left(\frac{\pi}{h^\vee} \langle \rho,\alpha\rangle\right)}\,.
\end{align}
This yields a more explicit expression for $\mathrm{I}_G^\lambda(q)$:
\begin{align}
\mathrm{I}_G^\lambda(q) = q^{\frac{c_2(\lambda)}{2 h^\vee}} \prod_{\alpha \in R_+} \frac{\sin\left(\frac{\pi}{h^\vee} \langle \lambda+ \rho,\alpha\rangle\right)}{\sin\left(\frac{\pi}{h^\vee} \langle \rho,\alpha\rangle\right)}\,.
\end{align}
   
We note that the coroot lattice form \eqref{latticeform} can be equally derived by a direct integration. The subtle point is that it will involve an orthogonal relation for non-dominant weight, which requires straightening, 
\begin{align} \label{ortho}
    \frac{1}{|W_G|} \oint_{T(G)} \frac{\mathrm{d}s}{(2 \pi i)^r s} w_H(z)\, \chi_{\gamma}(s)\, \chi_{\gamma'}(s^{-1}) = \delta_{\gamma, \gamma'}\, \quad \gamma,\gamma' \in h^\vee Q^\vee\,.
\end{align}
Such an orthogonality relation can, however, be justified by considering the shift Weyl action on the scaled coroot lattice $h^\vee Q^\vee$.
Note that the orthogonality relation \eqref{ortho} can be justified as follows: denoting the dot action $u \cdot \gamma = u (\gamma +\rho)-\rho$ for a Weyl group element $u$, then any $\gamma \in h^\vee Q^\vee$ admit a unique dominant representative $\lambda = u \cdot \gamma$ for a unique $u$, and such a correspondence is also injective. The injectivity follows from a key property of the Weyl vector $\rho$: for any nontrivial $w \in W$, $\rho - w(\rho) \notin  h^\vee Q^\vee$. As if If $u \cdot a = v \cdot b$, then
$u(a+\rho) = v(b+\rho) \Rightarrow u(a) - v(b) = v(\rho) - u(\rho) \in M$.
This implies $u = v$, so $a = b$, a contradiction. 


Furthermore, equivalent expressions involving the affine Grassmannian form follow from $\varepsilon(c) ^2\equiv 1$: 
    \begin{align}
        I_G(p,q) &= \sum_{c \in W^0} (pq)^{\langle \Lambda_0 - c(\Lambda_0),\, \hat{\rho} \rangle} \\
        \mathbb{I}_G(q) &= \sum_{c \in W^0} q^{2 \langle \Lambda_0 - c(\Lambda_0),\, \hat{\rho}\rangle}
    \end{align}
This gives the generalization of the connection to the combinatorics of $N$ core partitions given in Eq.~\eqref{suNcore} in the $SU(N)$ case, to general $G$. 

Finally, we can consider the classical theta function form of the Macdonald identity. First for $I_G(p,q)$,
\begin{align}
    & \frac{1}{|W|} \oint_{T(G)} \frac{\mathrm{d}s}{(2 \pi i)^r s} \,  J_{\hat{\rho}}(s, q) J_{\hat{\rho}}(s^{-1}, p)\\
 =& \, \frac{1}{|W|} \sum_{w, w'\in W} \varepsilon(ww') \sum_{\gamma,\gamma' \in Q^\vee} q^{\frac{|\rho + h^\vee \gamma|^2- |\rho|^2}{2 h^\vee}} p^{\frac{|\rho + h^\vee \gamma'|^2 - |\rho|^2}{2 h^\vee}} \oint \frac{ds}{(2\pi i)^r s} s^{w(\rho + h^\vee \gamma) - w'(\rho + h^\vee \gamma')} \\
  =& \, \frac{1}{|W|} \sum_{u, w'\in W} \varepsilon(u) \sum_{\gamma, \gamma' \in Q^\vee} q^{\frac{|\rho + h^\vee \gamma|^2 - |\rho|^2}{2 h^\vee}} p^{\frac{|\rho + h^\vee \gamma'|^2 - |\rho|^2}{2 h^\vee}} \delta_{0,\, u\cdot \gamma- \gamma'}\\ 
 =& \,   \sum_{\gamma \in Q^\vee} (qp)^{\frac{|\rho + h^\vee \gamma|^2 - |\rho|^2}{2 h^\vee}} \,.
\end{align}
where the condition $u \cdot \gamma-\gamma' =0$ is satisfied if and only if $u=e \in W$ and $\gamma' = \gamma$.
Hence, we get
\begin{align}
    I_G(p,q) = (pq)^{-\frac{|\rho|^2}{2h^\vee}}  \Theta_{\rho, h^\vee}(pq)\,. 
\end{align}
So it is a specialization of the classical theta function at $s=1$\,. 
Similar results hold for Schur index, 
\begin{align}
    \mathbb{I}_G(q)= q^{-\frac{|\rho|^2}{h^\vee}}  \Theta_{\rho, h^\vee}(q^2)\,.
\end{align}
Incidentally, Schur index is identified with the $p=q$ specialization of the superconformal 4d $\mathcal{N}=1$ SYM, 
\begin{align}
    I_G(q,q) = \mathbb{I}_G(q)\,. 
\end{align}
Remark that the following interesting relation holds:
\begin{align}
    J_{\rho}(a)^{-1} \sum_{w \in W}  \varepsilon(w) \Theta_{w(\hat{\rho})}(a,q) = \Theta_{\hat{\rho}}(q)\,.
\end{align}
which follows from 
\begin{align}
    \Theta_{w(\hat{\rho})}(a,q) = \Theta_{\hat{\rho}}(q) a^{w(\rho)}\,, \quad J_{\rho}(a) = \sum_{w\in W} \varepsilon(w) a^{w(\rho)}\,.
\end{align}

We can also similarly evaluate the Neumann half-index using the $W$-skew symmetric functions: 
\begin{align}
  &   \frac{1}{|W|} \oint \frac{ds}{(2\pi i)^r s}  \hat{J}_{\rho,h^\vee}(s,q) J_{\lambda+\rho}(s^{-1}) \\
  =& \, \frac{1}{|W|} \sum_{w, w'\in W} \varepsilon(ww') \sum_{\gamma \in Q^\vee} q^{\frac{|\rho + h^\vee \gamma|^2- |\rho|^2}{2 h^\vee}} \oint \frac{ds}{(2\pi i)^r s} s^{w(\rho + h^\vee \gamma) - w'(\lambda+ \rho)} \\
 = &  \, \frac{1}{|W|} \sum_{u, w'\in W} \varepsilon(u) \sum_{\gamma \in Q^\vee} q^{\frac{|\rho + h^\vee \gamma|^2 - |\rho|^2}{2 h^\vee}} \delta_{0,\,u\cdot(h^\vee \gamma)-\lambda}\\ 
 =& \, \varepsilon(u_\lambda ) q^{\frac{|\lambda+ \rho|^2 - |\rho|^2}{2h^\vee}} = \chi_{\lambda}(a) q^{\frac{|\lambda+ \rho|^2 - |\rho|^2}{2h^\vee}} 
\end{align}
where $u_\lambda$ satisfying the condition  $u\cdot ( h^\vee \gamma) = \lambda$ uniquely. 
This is identical to the previous results in Eq.~\eqref{defecthalfindex}.

Based on the explicit evaluations above, we further note the following bilinear product relation:
\begin{align}
    I_G(p,q) & =\, \sum_{\lambda \in P_{+}} \mathrm{I}_G^\lambda(p)\mathrm{I}_G^\lambda(q)\,, \label{n=1shurbilinear} \\
    \mathbb{I}_G(q) & = \, \sum_{\lambda \in P_{+}}\mathrm{I}_G^\lambda(q)^2\,. \label{n=2shurbilinear}
\end{align}
In fact, such relations can be established using the reproducing kernel for $G$:
\begin{align}
    K(s,t) = \sum_{\lambda \in P_+} \chi_{\lambda}(s) \chi_\lambda(t^{-1})\,,
\end{align}
which satisfies
\begin{align}
f(s)=  \frac{1}{|W|} \oint \frac{dt}{(2\pi i)^r t}w_H(t)K(s,t)f(t)
\end{align}
for any symmetric function 
$f$ under $W$. Then we may apply this to a Macdonald denominator $\mathcal{M}(s,q)$ in the integrand. We may apply this identity to a Macdonald denominator $\mathcal{M}(s,q)$
in the integrand. Then one obtains
\begin{align}
   I_G(p,q) 
   &= \frac{1}{|W|}
      \oint \frac{ds}{(2\pi i)^r s}\,
      \omega_H(s)\, \mathcal{M}(s,q)\,
      \Biggl(
        \frac{1}{|W|}
        \oint \frac{dt}{(2\pi i)^r t}\,
        \omega_H(t)\, K(s,t)\, \mathcal{M}(t,q)
      \Biggr) \\
   &= \sum_{\lambda \in P_+} 
      \mathrm{I}_G^\lambda(p)\, \mathrm{I}_G^\lambda(q)\,,
\end{align}
and the same argument applies to $\mathbb{I}_G(q)$.

\subsection{Interpretation}
\paragraph{Class $\mathcal{S}$ and TQFT wave functions}
First, let us consider the $4d$ $\mathcal{N}=2$ case. The $4d$ $\mathcal{N}=2$ SYM theories can be realized in terms of class $\mathcal{S}$ constructions \cite{gaiottoMN09}. For gauge groups of type $ADE$, the construction involves Higgs bundles and the Hitchin equations on a Riemann sphere with two identical punctures carrying specific irregular singularities. For the $BCFG$ cases, one expects a similar structure, though the analysis requires incorporating outer automorphism twists connecting the two singularities. In this framework, their $K$-theoretic Coulomb branch is identified with the Betti form of the Hitchin moduli space.

We find that the half-index $\mathrm{I}_G^\lambda(q)$ can be identified with the TQFT wavefunction associated with an irregular puncture in the class $\mathcal{S}$ construction of $4d$ $\mathcal{N}=2$ theories. This identification holds, at least, for the $ADE$ cases. Moreover, the bilinear form \eqref{n=2shurbilinear} involving $\mathbb{I}_G(q)$ reproduces the expected TQFT structure. Since this bilinear sum is related to the second form of the Macdonald identity, expressed in terms of specialized central heat kernels, it may further be interpreted as the cylinder partition function of 2d Yang–Mills theory with holonomies given by Kostant’s principal element of type $\rho$. On the other hand, the TQFT associated with class $\mathcal{S}$ involves the $q$-deformed 2d Yang–Mills theory \cite{gaddeRRY11}, and it would be interesting to clarify the precise relation between these two occurrences of 2d Yang–Mills theories.

For the $4d$ $\mathcal{N}=1$ case, there is likewise a class $\mathcal{S}$ construction involving irregular punctures. In this setting, one obtains the mixed Schur index \cite{beemG12}, which is again expressed as a bilinear form in the TQFT wavefunctions associated to the irregular punctures. The bilinear pairing \eqref{n=1shurbilinear} for $I_G(p,q)$ agrees with the TQFT results as well \cite{song15}.

\paragraph{Holomorphic twist, Affine Grassmannian and Schubert cells} 

Recall that the holomorphic twist and the holomorphic–topological twist can be used 
to define the $4d$ $\mathcal{N}=1$ and $\mathcal{N}=2$ indices, respectively, which 
count cohomology classes of operators with respect to twisted scalar nilpotent 
supercharges. This setup naturally extends to a broader framework of categorification 
with higher categories, incorporating both surface and line operators.\footnote{See 
\cite{kapustinSV10} for a discussion of the 2-category of surface defects in GL-twisted 
$4d$ $\mathcal{N}=4$ SYM theories.} In this categorical perspective, the spectrum of 
operators filtered by spacetime dimension is encoded in morphism spaces of increasing 
levels, arising from junctions of lower-codimensional objects. For instance, surface 
defects may be regarded as objects, while line operators—realized as junctions between 
surface defects—constitute the morphism spaces between them. An analogous relationship 
holds between line operators and local operators.

Let us first focus on the $4d$ $\mathcal{N}=2$ case with the holomorphic–topological 
twist on $\mathbb{C} \times \mathbb{R}^2$. Operators extending along $\mathbb{R}^2$ 
are topological and organize into a higher category consisting of surface operators, 
line operators, and local operators. Important examples arise from those supported 
at points of $\mathbb{C}$. Taking into account a natural $\mathbb{C}^\ast$ action 
on $\mathbb{C}$, one encounters cases where the supports are localized at the origin 
of $\mathbb{C}$, distributed along circles surrounding the origin, or extended over 
the entire $\mathbb{C}$-plane, all stable under the $\mathbb{C}^\ast$ action. 
With respect to the $\mathbb{C}^\ast$ action, the $\mathbb{C}$-plane can be deformed 
into a cigar. Further reduced on the cigar circle, it corresponds to a topological 
B-twisted $3d$ $\mathcal{N}=4$ theory with a boundary corresponding to the origin of 
$\mathbb{C}$. By this, the 2-category of surface operators $\mathbb{S}$ maps to the higher category of boundary 
conditions in the 3d topological B-twisted theories\footnote{The 2-category of 
surface operators in 4d allows for braiding, while the 2-category of boundary conditions 
in 3d only allows for fusion.}. 

The identity object in the 2-category $\mathbb{S}$ is the trivial surface operator 
$S_{id}$, and bulk line operators belong to its endomorphism space, forming a category. 
The category with bulk line operators as objects and defect-changing local operators 
as morphisms is of great interest in the $4d$ $\mathcal{N}=2$ setting, due to the 
close connection between line operators, Coulomb branch geometry, and the spectrum of 
BPS particles.

For $4d$ $\mathcal{N}=2$ gauge theories (including pure Yang–Mills theories) with 
gauge group $G$, the corresponding description is provided by the $B$-twist of the 
$3d$ $\mathcal{N}=4$ gauge theory whose gauge group is the loop group $LG$. Hence, 
there are more geometric formulations in terms of the loop group $LG$. In particular, 
for $\mathcal{N}=2$ SYM theories, the category of bulk line operators is given by
\begin{align}
\mathbb{L} = \mathrm{End}_{\mathbb{S}}(S_{id}) = D^{b}\mathrm{Coh}_{G_{\mathbb{C}}[[z]]}
\big(\mathrm{Gr}(\hat{G})\big),
\end{align}
known as the \emph{coherent Satake category}~\cite{cautisW19}. Here, 
$\mathrm{Gr}(\hat{G})$ denotes the affine Grassmannian, the quotient of the loop group 
$G_{\mathbb{C}}((z))$ by the subgroup $G_{\mathbb{C}}[[z]]$ acting by left multiplication:
\begin{align}
\mathrm{Gr}(\hat{G}) = G_{\mathbb{C}}((z))/G_{\mathbb{C}}[[z]]\,.
\end{align}

The affine Grassmannian admits a Schubert cell decomposition with strata given by 
the orbits of the Iwahori subgroup~\cite{lamLMSSZ14}. These cells are labeled by 
elements of the coroot lattice of the affine Weyl group. Although the coherent Satake 
category $\mathbb{L}$ is not semi-simple, it still possesses a well-defined ring of 
simple objects~\cite{cautisW19}. A distinguished collection of simple objects is 
realized by coherent sheaves supported on Schubert cells.

The Schur index can then be understood as the trace of charge operators on the space 
of bulk local operators,
\begin{align}
\mathcal{V} = \mathrm{End}_{\mathbb{L}}(L_{id}),
\end{align}
so that it is determined by the eigenvalues of the charge operators acting on a 
basis of simple line operators. Our results are consistent with this categorical 
picture: in both descriptions, the index is expressed as a sum over the coroot lattice 
(or equivalent parametrizations).

We could also consider the setting with boundaries/interfaces, which are supported 
on the product of $\mathbb{C}$ and a 1d locus on the topological $\mathbb{R}^2$. 
In particular, the half Schur index of $4d$ $\mathcal{N}=2$ SYM theories with Neumann 
boundary conditions, defined either as a Witten index on $HS^3 \times \mathbb{R}$ 
or via HT twist on $\mathbb{C} \times \mathbb{R} \times \mathbb{R}_+$, decorated by 
BPS Wilson line insertions, play a central role. It counts the local operators 
on an interface between the empty theory and $4d$ $\mathcal{N}=2$ SYM.\footnote{
In the holomorphic-topological twisting setting, the interface may give rise to certain 
line defects along the topological $\mathbb{R}^2$, by reduction on the holomorphic 
$\mathbb{C}$. The holomorphic $\mathbb{C}$ dependence introduces a $q$-grading 
from the rotation on $\mathbb{C}$.}

For $\mathcal{N}=1$ gauge theories, the cigar reduction leads to 3d 
$\mathcal{N}=2$ loop group $LG$ gauge theories with holomorphic-topological twists. 
The affine Grassmannian also arises naturally in this context~\cite{costelloDG20,garnerW23}.

\paragraph{IR formula and BPS spectrum}

Another important piece of IR physics related to Seiberg-Witten solution is the BPS spectrum, which can be captured by a BPS quiver $Q$ with superpotential $W$ and cohomological Hall algebras (CoHA) $\mathrm{H}(Q,W)$ \cite{gaiottoGL24}.  CoHA also exhibits cluster algebra structure, which is related to the cluster algebras of line operators through the Coulomb branch RG flow \cite{gaiottoGL24}. Furthermore, it is related to the wall-crossing behavior of the BPS spectrum. Recall that the BPS spectrum is known to exhibit wall-crossing behavior on the Coulomb branch, corresponding to cluster transformations. An important type of wall crossing invariant is given by the KS quantum spectrum generators, which is given by the character of CoHA. There is an IR formula that relates the trace of the KS quantum spectrum generator and its double to the half index and Schur index, respectively. Our exact formula gives a concrete relation as the trace of the double of the KS operator as a bilinear sum of traces of the KS operators. 

The BPS monodromy operator admits a factorization
\begin{align}
    \mathcal{K}(q) = \mathcal{S}(q) \mathcal{S}^{t}(q)
\end{align}
where $\mathcal{S}(q)$ is the Kontsevich-Soibelman (KS) quantum spectrum generator is a formal power series in the quantum torus algebra and $q$.  
The inverse of $\mathcal{S}(q)$ can be identified with the character of CoHA:
\begin{align}
    \mathcal{S}^{-1}(q) = \chi(\mathrm{H}(Q,W))\,.
\end{align}

Then the IR expressions for the Schur index and the half index take the form
\begin{align}
\mathbb{I}_G(q) = (q;q)^{2r}\, \mathrm{Tr}\!\left[\mathcal{K}(q)\right],
\qquad
\mathrm{I}_G^{\lambda}(q) = (q;q)^{r}\, \mathrm{Tr}\!\left[\mathcal{S}(q)\, W_{\lambda}\right],
\end{align}
where $W_{\lambda}$ denotes the IR operator associated with the Wilson line in the irreducible representation of highest weight \(\lambda\).
It plays the role of the character of the corresponding highest–weight module.

By the relation between the Schur index and the half index, we also have a similar relation for the IR formulas as follows
\begin{align} \label{bilinIR}
    \mathrm{Tr}[\mathcal{S}(q) \mathcal{S}^{t}(q)] = \sum_{\lambda} \left( \mathrm{Tr}[\mathcal{S}(q) W_\lambda ]  \right )  \left( \mathrm{Tr}[\mathcal{S}(q) W_\lambda ] \right)
\end{align}

On the UV side, the integration amounts to extracting the constant term by Weyl integration, whereas on the IR side it corresponds to a trace operation given by the projection $X_\gamma \to \delta_{\gamma,0}$. 

There is a structural parallel between the UV and IR descriptions. The half BPS monodromy multiplied by $(q,q)^r$ corresponds to the Macdonald–Kac denominator, while the full BPS monodromy multiplied by $(q,q)^{2r}$ corresponds to the square of the Macdonald–Kac denominator. The trace operation corresponds to the Haar integration. On the IR side, we work with non-commutative quantum torus variables, whereas on the UV side the variables live on the Cartan torus of $G$.
The bilinear decomposition of the Schur index in terms of the half-index with Wilson lines thus translates into an operation on the trace of the full BPS monodromy.
On the integration side, one can insert a reproducing kernel; analogously, there should exist a corresponding operation at the level of the quantum torus and the BPS monodromy. This suggests the schematic relation
\begin{align}
    \mathrm{Tr}[\mathcal{S}(X) \mathcal{S}^t(X)] \;=\; \mathrm{Tr}_X\Big[\mathcal{S}[X] \, \mathrm{Tr}_Y[C(X,Y) \mathcal{S}^t(Y)]\Big],
\end{align}
where the inner trace acts on the $Y$-variables, and the outer trace on the $X$-variables. Here 
\begin{align}
    C(X,Y) \;=\; \sum_\lambda W_{\lambda}(X) W_{\lambda}^t(Y)
\end{align}
serves as an analogue of the reproducing kernel. Then we find
\begin{align}
    \sum_\lambda \mathrm{Tr}_X[\mathcal{S}[X] \, \mathrm{Tr}_Y[W_\lambda(X) W_\lambda^t(Y)\mathcal{S}^t(Y)]]
= &  \sum_\lambda \mathrm{Tr}_X[\mathcal{S}[X] W_\lambda(X)] \; \mathrm{Tr}_Y[W_\lambda^t(Y)\mathcal{S}^t(Y)] \\ 
= &  \sum_\lambda \mathrm{Tr}_X[\mathcal{S}[X]W_\lambda(X)] \; \mathrm{Tr}_Y[\mathcal{S}(Y)W_\lambda(Y)]\,.
\end{align}
which exactly reproduces Eq.~\eqref{bilinIR}. 

\section{Conclusion and Outlook}
\label{conclusion}

In this work we have established closed-form evaluations of the supersymmetric index for 4d $\mathcal{N}=1$ and $\mathcal{N}=2$ Super Yang-Mills theories with an arbitrary simple gauge group. The key tool was the use of Macdonald identities for affine Lie algebras, which allowed us to reformulate the index in several equivalent ways: in terms of the Kostant–Fegan form, as a lattice–sum expression, through affine Grassmannian forms, and also with classical theta series. A further application of specialized Macdonald identities yields product formulas in terms of eta–quotients, exhibiting interesting modular properties. We also derived exact expressions for the half-Schur index with Neumann boundary conditions and Wilson lines, and identified a bilinear sum representation of the full index.

Beyond their technical derivation, our results admit a natural interpretation within the framework of holomorphic and holomorphic–topological twists. In particular, for 4d $\mathcal{N}=2$ theories, this opens the avenue toward a categorification in terms of the category of BPS line operators. This perspective further suggests a broader framework linking supersymmetric indices, Schur quantization, and the BPS spectrum.

Finally, our analysis points to several promising directions. Extensions to twisted affine algebras and to super affine Lie algebras appear especially promising, and some progress in these directions will be reported elsewhere. Moreover, the relation to quantum integrable systems for the $K$-theoretic Coulomb branch deserves further exploration. Furthermore, there is an interesting connection between real Schur quantization and the double-scaled SYK model \cite{gaiottoV24, lewisMSS25, berkoozKS25}. It would also be interesting to explore non-conformal holography and the gravity dual of the superconformal index of super Yang-Mills theories using various tools including giant gravitons, non-critical strings \cite{deiFR24}. Altogether, we expect these developments to enrich both the physical and mathematical understanding of supersymmetric gauge theories.

\section*{Acknowledgments}

We are grateful to O.~Chalykh, S.~Kato, K.~Maruyoshi, T.~Okazaki,
Y.~Pan, M.~Shimozono, D.~Wahiche, O.~Warnaar, and X.~Yin 
for valuable discussions and helpful comments.  
This work was initiated during the ``Kyushu IAS–iTHEMS Workshop: Non-perturbative Methods in QFT"
and has been presented at the ``YITP Workshop: New Advancements on Defects and Their Applications,"
whose stimulating environments are gratefully acknowledged. This work is supported by the KIAS Individual Grant PG084801.

\bibliographystyle{ytamsalpha}
\bibliography{mybib}

\end{document}